\documentclass[reqno]{amsart}
\usepackage{graphicx}
\usepackage{amsmath}
\usepackage{amssymb}
\usepackage{amsthm}
\usepackage{latexsym} 
\usepackage{thumbpdf}
\usepackage{subfigure}
\usepackage{float}
\usepackage[table]{xcolor}
\usepackage{url}
\usepackage{array}
\usepackage{bigstrut}
 \usepackage{colortab}
\usepackage{pstricks}
\usepackage{pst-node}

\usepackage{tikz}
\usetikzlibrary{arrows}
\usetikzlibrary{matrix}

\def\ds{\displaystyle}

 \title[How Infectious Was Deflategate?]{How Infectious Was \#Deflategate?}

 \author{Eric Eager$^{\rm a,b}$, Megan Eberle$^{\rm a}$, and James Peirce$^{\rm a,b}$}
  \address{  
  University of Wisconsin - La Crosse}
  
  \address{$^{\rm a}${\em{Mathematics Department, University of Wisconsin - La Crosse}}, \\
$^{\rm b}${\em{River Studies Center}}}

\begin{document}

\begin{abstract}
On Monday January 19, 2015 a story broke that the National Football League (NFL) had started an investigation into whether the New England Patriots deliberately deflated the footballs they used during their championship win over the Indianapolis Colts.  Like an infectious disease, discussion regarding Deflategate grew rapidly on social media sites in the hours and days after the release of the story.  However, after the Super Bowl was over, the scandal slowly began to dissipate and lost much of the attention it had originally had, as interest in the NFL wained at the completion of its season. We construct a simple epidemic model for the infectiousness of the Deflategate news story. We then use data from the social media site Twitter to estimate the parameters of this model using standard techniques from the study of inverse problems.  We find that the infectiousness (as measured by the basic reproduction number $\ds \mathcal{R}_0$) of Deflategate rivals that of any infectious disease that we are aware of, and is actually more infectious than recent news stories of greater importance - both in terms of $\ds \mathcal{R}_0$ and in terms of the average amount of time the average tweeter continued to tweet about the news story. 
\end{abstract}

\maketitle

  \section{Introduction}

The National Football League (NFL) has made American football arguably the most popular sport throughout the United States.  The NFL was formed in 1922 from the American Professional Football Association, and originally consisted of 18 teams.  By 1925 the league began drawing tens of thousands of fans into stadiums to watch games live, and by  1934 the first NFL game between the Chicago Bears and the Detroit Lions on Thanksgiving Day was broadcasted live on national radio - allowing the league's fan base to spread nationally.  The NFL's popularity continued to grow when the Brooklyn Dodgers played the Philadelphia Eagles was the first NFL game to be televised on the National Broadcasting Company(NBC).  Numerous other professional football leagues were formed to compete with the NFL during this time - although none could compete against the NFL until the All-America Football Conference (AFC), which later merged with the NFL in 1949 to further increase the size and fan base of the NFL.  As the radio and television became more widespread across the United States, the NFL was able to increase their popularity as almost all NFL games were broadcasted on radio or television by 1964, allowing professional football to overtake professional baseball in popularity around 1965.  As the years progressed, the popularity of the NFL continued to grow as television ratings surged into the millions viewers and stadiums grew to house over 100,000 fans.  Today, the fan base of the NFL well over 100 million people, with increasing popularity overseas \cite{NFL}.

Despite the popularity of the NFL, its history has been plagued by numerous scandals; most recently, a scandal involving the footballs used by the New England Patriots in the 2015 AFC Championship game against the Indianapolis Colts, which has become known as ``Deflategate".  On Monday January 19, 2015 a story broke that the NFL had started an investigation into whether the New England Patriots deliberately deflated the footballs they used during their AFC Championship win over the Indianapolis Colts.  The NFL rules state that footballs must weigh between 14 and 15 ounces and be inflated between 12.5 and 13.5 pounds per square inch.  Deflated footballs would have given the Patriots an advantage over the Colts (as well as other opponents) by providing a better grip on the ball \cite{Clark}.  

Discussion was sparked instantly on social media regarding Deflategate, and grew rapidly in the hours and days after the release of the story. This scandal gained national attention quickly as the New England Patriots had just earned a trip to Super Bowl XLIX against the Seattle Seahawks (another team displaying arguable ethics of its own in previous seasons \cite{PEDs}).  Since the Super Bowl is the most anticipated game in the NFL season, attention to the story was heightened. After the Super Bowl was over, the scandal slowly began to dissipate and lost much of the attention it had originally had, as interest in the NFL decreased at the completion of its season.This interest continued to dissipate until early May, 2015, when the aftermath of a 243-page report by independent attorney Theodore V. Wells, Jr. resulted in the NFL suspending Patriots star quarterback Tom Brady for the first four games of the 2015 season and stripping the team of two high draft picks.  

The sharp rise in interest in the Deflategate scandal and then an initial slow dissipation of interest is similar to that of an outbreak of an infectious disease.  When an infectious disease is first introduced to a new population, it can spread rapidly as a large number of the population becomes infected with the disease.   The spread of the disease slows to more manageable rate as more of the population becomes immune to the disease through vaccinations or survival of victims of the disease, leaving less of the population susceptible to the disease.  The infectiousness of a disease can often be captured by a dimensionless parameter called the \textit{basic reproduction number} $\ds \mathcal{R}_0$ (``R-naught").  The value of $\ds \mathcal{R}_0$ informs scientics of how fast the disease is initially spreading by indicating the number of individuals an initial infected individual is likely to infect.  For example, the $\ds \mathcal{R}_0$ value for the Ebola virus was 2, meaning that if one person was infected with Ebola during the early stages of the epidemic, they would likely spread the virus to 2 other people before they recovered or passed away.  Some other $\ds \mathcal{R}_0$ values include 4 for the HIV and SARS virus, 10 for mumps and 18 for measles \cite{Doucleff}.  An $\ds \mathcal{R}_0$ value of 18 is extremely high and means that measles was extremely infectious; which is why it was so catastrophic.  

Since the Deflategate scandal spread in a way similar to an infectious disease, we can derive and use something like an $\ds \mathcal{R}_0$ value for this news event in order to understand how infectious this story was.  To do this we must determine a useful medium through which information was spread, which we determined to be the social networking site Twitter (http://twitter.com).  Twitter allows users post \textit{tweets}, a message under 160 characters, to their Twitter page so that their followers, other users who will see the tweet, can read their message.  The followers can then reply to the message and start a conversation with the owner of the tweet.  They can also simply \textit{retweet} that tweet to their followers.  A retweet is when follower takes a tweet they saw and posts it onto their Twitter page for their followers to see.  The number of tweets and retweets about Deflategate can be used to determine the $\ds \mathcal{R}_0$ value and the infectiousness of this scandal as a news story. 

In this paper we construct a simple SIR epidemic model for the infectiousness of the NFL's Deflategate news story.  We use standard techniques from epidemiology to derive the basic reproduction number $\mathcal{R}_0$ for this model, as well as the average amount of time the average individual tweets about the story.  We then use Twitter data to estimate the parameters of this model using standard techniques from the study of inverse problems.  We find that the infectiousness (as measured by $\mathcal{R}_0$) of Deflategate rivals that of any infectious disease that we are aware of, and is actually more infectious than the story of Hillary Clinton's announcement of her presidential campaign in April, 2015 - both in terms of $\mathcal{R}_0$ and in terms of the average amount of time the average tweeter continued to tweet about the news story.  We also show that the average individual tweeted about Deflategate roughly ten times more than they did about the story of Freddie Gray's death that elicited the Baltimore riots in April of 2015.  

 \section{The Model}
 
While the punishment of the Patriots in May, 2015 created its own news story regarding Deflategate, we will consider only the dynamics of the initial story, occuring in January, 2015.  We assume that information concerning Deflategate is spread person to person like an acute infection.  Since online discussion of specific topics is often measurable, we will focus on the subpopulation of media consumers that use the social media site Twitter.   Time $t$ will be measured in days since the first report of Deflategate on January 19th, 2015. Using formalism first studied in depth by Kermack and McKendrick \cite{Kermack1927}, and popularized by Brauer and Castillo-Chavez \cite{CCC}, to name just two, we categorize the individuals in the Twitter population as susceptible, infectious, and removed. The \textit{susceptible population} $S(t)$ (measured in thousands) are those individuals that regularly read and comment on sports news but have yet to comment on Deflategate as of time $t$.  The \textit{infectious population} $I(t)$ (measured in thousands) are those that are tweeting (or retweeting) posts about Deflategate using the keywords \#deflategate, deflate gate, deflate-gate, spygate, or ``deflated balls."  As a simplifying assumption we consider each tweet as representing a unique individual (tweeter), and thus tweets and individuals in the infectious class are discussed interchangably.  The \textit{removed population} $R(t)$(measured in thousands) includes either those individuals who do not read or comment on sports (and are hence removed from this news event even from the very beginning), or individuals that were once tweeting about the story and have permanently stopped using the Deflategate keywords.  Due to the short lifespan of the story (in the order of weeks), we assume that the immigration of new users, emigration of current uses, birth, and death can be neglected, yielding a constant total population size. 

 \begin{figure}[h!]
    \begin{center}
\begin{tikzpicture}[->,>=stealth',shorten >=1pt,auto,node distance=3cm,
  thick,main node/.style={rectangle,draw,font=\sffamily\Large\bfseries}]

\node[main node] (S) at (0,0) {$S$};
\node[main node] (I) at (3,0) {$I$};
\node[main node] (R) at (6,0) {$R$};

\node[] (SI) at (1.5,0) {};

\path[->]
	(S) edge node {} (I)
	(I) edge[dashed, bend right=60] node[anchor=north, above] {$\beta$} (SI)
	(I) edge node[above] {$\gamma$} (R); 
	
\end{tikzpicture}
 \caption{Conceptual model of information transmission and recovery into the removed population. Black arrows show the movement between the $S$ and $I$ classes and the $I$ and $R$ classes. The fact that the level of the size of the infection population influences the rate at which a susceptible individual moves in to the infectious class is show by the dashed arrow.} \label{fig:concept}
 \end{center} 
 \end{figure}
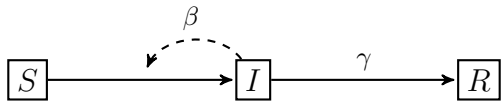

The progression from the susceptible population $S$ to the infectious population $I$ to the recovered (or - maybe more appropriately named - bored) population $R$ can be visualized in the traditional SIR conceptual diagram (Figure \ref{fig:concept}).  The progression of the information through the population depends on many factors. One of the most prominent is the total number of ``interactions" 
between susceptible and infectious individuals.  Information about Deflategate spreads when a susceptible 
individual comes in contact with the information spread by an infectious individual (by reading his or her tweets) and subsequently 
becomes infectious (starts tweeting themselves).  Mathematically, a reasonable measure for the number of encounters 
between susceptible and an infectious individuals, assuming 
homogeneous mixing, is given by the product $S I$.  This is referred to as the \textit{law of mass action} in the applied mathematics literature \cite{Holmes2009}.  However, not every interaction of a susceptible person reading a tweet about Deflategate will cause a retweet or a series of original tweets by the newly-infectious.  We 
use a parameter $\beta$, the transmission coefficient, as a daily measure of the probability that a single susceptible reader will retweet using the keywords \#deflategate, ``deflate gate", ``deflate-gate", ``spygate" (a previous Patriots scandal), or ``deflated balls" upon encountering such tweets in his or her feed.   Other functional forms exist to model the rate of new infections (for example, the Holling Type II functional form \cite{Holling1959}), which we will also explore.  Once infectious, we assume that individuals tweets about this story for an average of $1/\gamma$ days. The following assumptions can be collected to create the following SIR model
 \begin{align}
 \begin{split}
 S'(t) &= - \beta S(t) I(t) \\
 I'(t) &= \beta S(t) I(t)  - \gamma I(t) \\
S(0) &= S_0, \; \; I(0) = I_0, 
 \end{split}\label{eq:sir}
 \end{align}
where $S_0$ and $I_0$ are parameters giving the initial population of susceptible and infectious individuals (measured in thousands).  Since the $R(t)$ population doesn't interact with the rest of the population at any given time, we can omit it from our analysis. We assume that the parameters $\beta, \gamma, S_0$ and $I_0$ are all positive, which implies that the solution vector $[S(t) \; \; I(t)]^T$ exists and is positive for all $t > 0$ \cite{SmithThieme2011}.   

The derivation of the basic reproduction number $\ds \mathcal{R}_0$ for this SIR model follows from linearizing about the ``disease-free" state $[S_0 \; \; 0]^T$, where the susceptible population has yet to be exposed to the news story.  $\ds \mathcal{R}_0$ will represent the number of susceptibles that are initially ``infected" by an individual already tweeting about the story and whether the story will spread rapidly and receive attention or will receive little attention causing the story to fall from the headlines and die off.  

According to the SIR model \ref{eq:sir}, $S(t)$ will always be a decreasing function.  However for a small initial tweeting population $I_0$, the rate of change of $S$ maybe small and therefore we assume $\ds S(t) \approx S_0$.  Consequently, the equation for $I(t)$ reduces to  
\begin{equation}\label{eq:Ireduced}
I'(t) \approx \beta S_0 I(t)-\gamma I(t) = (\beta S_0-\gamma)I(t) = k I(t)
 \end{equation}
where $k$ is a constant that is equal to $\beta S_0-\gamma$.  The solution to the linear equation (\ref{eq:Ireduced}) is $\ds I(t) \approx I_0 e^{kt}$. The value of $k$ determines whether the function increases or decreases for small $t$.  The infected population will initially increase when $k = \beta S_0-\gamma > 0$.  Or equivalently,
\begin{align*}
\beta S_0-\gamma &> 0 \\
\frac{\beta S_0}{\gamma} &> 1
\end{align*}
Similarly the infected population will decrease if $k = \beta S_0/ \gamma <1$. This threshold is the basic reproduction number $\mathcal{R}_0=\beta S_0/ \gamma$.  One can interpret this value as the total rate of initial infections ($\beta S_0$) times the average amount of time spent infected $(\gamma^{-1})$.  Thus, for fixed $S_0$, diseases with large $\beta$ and/or small $\gamma$ will be the the most likely diseases to result in epidemics.  

\section{Methods}\label{section:method}
The size of the infected population during the days after Deflategate was first reports was gleaned from an article\cite{Cappadona2015} published in Boston Magazine. Data presented in the article was compiled by the website Topsy (http://www.topsy.com), an analytics company and certified Twitter partner, that collects Twitter data over the span of 30 days.  The number of tweets per day using the keywords \#deflategate, ``deflate gate", ``deflate-gate", ``spygate", or "deflated balls" is given by the unfilled circles in Figure 2.

We used this Twitter data to reverse-engineer the values of the parameters $\beta, \gamma, S_0$ and $I_0$ in the model (\ref{eq:sir}) by using standard methods from the study of inverse problems \cite{Banks2014}, which we summarize below.

To employ the techniques from inverse problems we require a \textit{statistical model} to go along with the mathematical model (\ref{eq:sir}).  To create such a statistical model, we abstract our mathematical model (\ref{eq:sir}) as in \cite{Avery2012} to give us 
\begin{eqnarray} \label{abstract}
\vec{x}'(t, \vec{\theta})’ &=& \vec{g}(\vec{x}(t, \vec{\theta}), \vec{\theta}), \; \; t \in [t_0, t_f],\\
\vec{x}(t_0, \vec{\theta})  &=& \vec{x}_0 \nonumber
\end{eqnarray}
where $\vec{x} (t, \vec{\theta})= [S(t),  \; I(t)]^T$ and $\vec{x}_0= [S_0,  \; I_0]^T$ are the vectors of state variables and initial conditions of our system, given the parameter vector $\vec{\theta} = [\beta,  \; \alpha,  \; S_0,  \; I_0]^T$.  We define as our \textit{observation process} the following
\begin{equation*}
f(t,\vec{\theta}) = C\vec{x}(t, \vec{\theta}) = I(t),
\end{equation*}
as we are only aware of the infectious population (i.e. the tweeters) at any given time $t$.  In this case $C = [0  \; 1 ]$ is a functional over $\mathbb{R}^2$ - the case where we were able to see the entire system $[S(t)\; \; I(t)]^T$ at any given time the operator $C$ would be the $2 \times 2$ identity matrix.  

To find an estimate for the parameter vector $\vec{\theta}$ we formulate the statistical model
\begin{eqnarray*}
Y(t) &=& f(t,\vec{\theta}_0) + \epsilon(t) \\
&=& I(t, \vec{\theta}_0) + \epsilon(t), \; \; t \in [t_0, t_f],
\end{eqnarray*} 
where $\vec{\theta}_0 = [\beta_0, \alpha_0, \; S_{00}, \; I_{00}]$ is considered the vector of hypothesized ``true values" of the unknown parameters and $\epsilon(t)$ is a random variable that represents observation error for the observed state variable at each time $t$.  We assume that the error function $\epsilon$ is such that $E(\epsilon(t)) = 0$, $Var(\epsilon(t)) = \sigma^2$ and $Cov(\epsilon(t)\epsilon(s)) = \sigma^2\delta(t - s)$ for all $t, s \in [t_0, t_f]$.  

Since our data is collected at discrete time points, we have to write our above statistical model in discrete terms.  Given data $I_1, I_2, ..., I_n$ taken at time points $t_0 \leq t_1 < t_2 < t_3 < ... < t_n \leq t_f$ we write our observation process as
\begin{equation*}
f(t_j,\vec{\theta}) = I(t_j, \vec{\theta}), \; \; j = 1, 2, ..., n,
\end{equation*}
with the discrete statistical model as
\begin{equation*}
Y_j = f(t_j,\vec{\theta}_0) + \epsilon(t_j) =  I(t_j, \vec{\theta}) + \epsilon(t_j)
\end{equation*}
and a realization of the discrete statistical model as
\begin{equation*}
I_j = f(t_j,\vec{\theta}_0) + \epsilon_j = I_j + \epsilon_j.
\end{equation*}
Knowing the value $\vec{\theta}$ and solving the system of differential equations (\ref{eq:sir}) with these parameter values is known as the \textit{forward problem}.  Alternatively, having a set of data $I_1, I_2, ..., I_n$ and estimating $\vec{\theta}_0$ is known as solving an \textit{inverse problem}.  In this paper we perform the latter.    

While several methods exist to solve inverse problems \cite{Avery2012}, we will use ordinary least squares.  In addition to the assumptions above, we assume that realizations of $\epsilon(t)$ at particular time points are independent and identically distributed normal random variables.  In this case, one can show that the parameter vector $\vec{\theta}_0$ that maximizes the likelihood function 
\begin{equation*}
\mathcal{L}(\vec{\theta}_0) = P(I_j = I(t_j, \vec{\theta}), j = 1, 2, ..., n|\vec{\theta} = \vec{\theta}_0) 
\end{equation*}
is given by the parameter vector that minimizes the least-squares functional
\begin{equation*}
\mathcal{LL}(\vec{\theta}_0) = \sum_{j = 1}^n(I_j - I(t_j, \vec{\theta}) )^2.
\end{equation*}
In other words
\begin{equation} \label{min}
\vec{\theta}_0 = \text{argmin}_{\vec{\theta} \in \Theta}\mathcal{LL}(\vec{\theta}),
\end{equation}
where $\Theta$ is the set of admissible values for the parameter vector $\vec{\theta}$.  For the model in this paper
\begin{equation*}
\Theta = (0, \infty)\times(0, \infty)\times(0, 3.02 \times 10^{8})\times(0, 3.02 \times 10^{8}),
\end{equation*}
where $3.02 \times 10^{8}$ are the estimated number of monthly Twitter users as of 5/14/2015 \cite{Twitter}.  

Once an estimate for parameter vector $\vec{\theta}_0$ is found, the next question is in regards to the uncertainty in the estimate.  To generate standard errors for each of the parameters we create $1 \times 4$ \textit{sensitivity matrix} 
\begin{equation*}
D_j(\vec{\theta}) = \left[\frac{\partial I (t_j, \vec{\theta})}{\partial \theta_1} \; \; \; \frac{\partial I (t_j, \vec{\theta})}{\partial \theta_2}  \; \; \; \frac{\partial I (t_j, \vec{\theta})}{\partial \theta_2}  \; \; \; \frac{\partial I (t_j, \vec{\theta})}{\partial \theta_2} \right],
\end{equation*}
from which we create the $4 \times 4$ convariance matrix
\begin{equation*}
\Sigma^n(\vec{\theta}_0) = (\sigma^2)^{-1}\left(\sum_{j=1}^nD_j^T(\vec{\theta}_0)D_j(\vec{\theta}_0)\right)^{-1}.
\end{equation*}
It follows that the standard error in the $k$th component of the parameter vector $\vec{\theta}_0$ can be approximated by the square root of the $k,k$th element of the matrix $\Sigma^n(\vec{\theta}_0)$ \cite{Avery2012}.  We report parameter values in an interval with the lower and upper bounds being two standard errors below and above the estimated value, respectively.  

All programming was performed in R \cite{R2012}, with code available upon request.  To solve this inverse problem we used a fourth-order Runge-Kutta \cite{Suli2003} to solve the forward problem (\ref{abstract}) for each set of parameters $\vec{\theta}$ in $\Theta$, and used the built-in optimization algorithm ``optim" in R \cite{R2012} to solve the minimization problem (\ref{min}) for $\vec{\theta}_0$.  

\section{Results}
For the data in Figure 2 a parameter vector that minimized the value of $\mathcal{LL}$ is given by $$\vec{\theta}_0 = [\beta_0 \;\; \gamma_0 \;\; S_{00} \;\; I_{00}]^T = [0.0153 \; \; 0.3643 \;\; 156.6120 \; \; 2.2723]^T.$$  Both and $S(t)$ and $I(t)$ components of the solution to (\ref{eq:sir}) subject to these parameter values is displayed in Table 1, along with the Twitter data used to solve the inverse problem.  95-percent confidence intervals for the parameters are given in Table 1.  Since all of these confidence intervals exclude zero, we can be fairly certain that each of the parameters in the model are significantly different than zero, and thus necessary to include.  While there were other parameter values that were (local) minimums of $\mathcal{LL}$ in the space $\Theta$, the fit in Figure 2 suggests that $\vec{\theta}_0$ is a reasonable estimate for the parameter vector $\vec{\theta}$. 

The parameter values in $\vec{\theta}_0$ are able to give us quantitative information regarding the infectiousness of the Deflategate news story.  For example, the estimate for $\beta$ suggests that between 1.3 and 1.7 percent of susceptible tweeters' views of Deflategate tweets will result in the immediate conversion of said tweeter into someone tweeting about the Deflategate story.   The estimate for $\gamma$ suggests that the average Deflategate tweeter spends between 2.24 and 3.53 days tweeting about the Deflategate story before becoming bored with the story.  

Since our units for $S$ and $I$ are in terms of thousands of tweeters, our value for the basic reproduction number $\mathcal{R}_0 = \beta S_0 (\gamma)^{-1}$ actually corresponds to the average number, \textit{in thousands}, of new Deflategate tweeters elicited by the average initial thousand Deflategate tweeters.  Our estimates in $\vec{\theta}_0$ therefore suggest a basic reproduction number between 4.05 and 10.91 thousands of new Deflategate tweeters per initial thousand Deflategate tweeters, a number rivalling some of the aforementioned epidemics in history.

 \begin{table} 
\begin{center}
\begin{tabular}{|c|c|c |c|} \hline
Parameter & lower bound & estimate & upper bound \\ \hline
$\beta$ & 0.0130 & 0.0153 & 0.0177 \\ \hline
$\gamma$ & 0.2830 & 0.3643& 0.4456 \\ \hline
$S_0$ & 139.0034 & 156.6120& 174.2207 \\ \hline
$I_0$ & 0.6626 & 2.2726&  3.8827 \\ \hline
  \end{tabular}
\end{center}
\caption{Parameter values and their respective 95\% confidence intervals.}
\end{table} \label{table1}

\begin{figure} 
\begin{center}
\includegraphics[scale = .5]{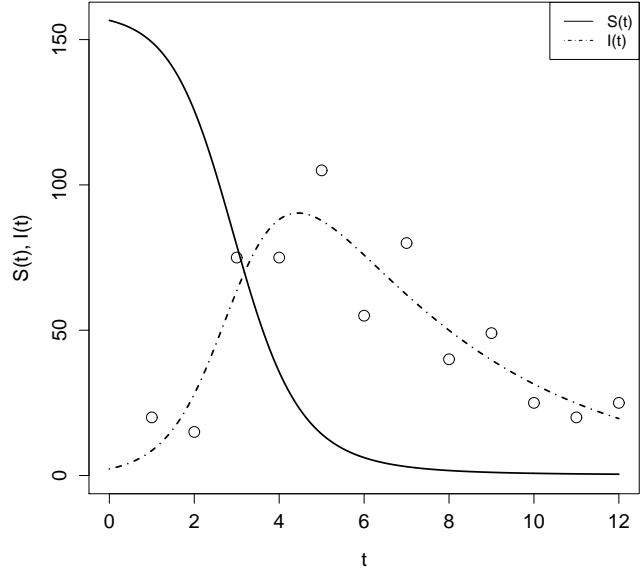}
\end{center}
\caption{The number (in thousands) of Twitter users susceptible, $S(t)$, and tweeting about Deflategate, $I(t)$, subject to parameter vector $\vec{\theta}$ solving the inverse problem using Topsy.com data (unfilled circles). $t=0$ corresponds to January 17, 2015.}
 \label{fig2}
\end{figure}

When viewing $S$ and $I$ together on the same set of axes we see that by the sixth day of the news story the entire population has either become a Deflategate tweeter themselves or was never going to tweet about Deflategate in the first place.  On the other hand, the values of $I$ continue to be well above zero almost two weeks into the news story.  These two pieces of information suggest that this news story - and possibly NFL news stories in general - are quite infectious and have a relatively high amount of staying power, which coincides with our initial intuition \cite{NFL}.

We also explored the possibility of another functional form for the probability of an ``infection", using the Holling type II \cite{Holling1959} functional form $$H(S, I) = \frac{\beta S I}{1 + hS}$$ in place of traditional mass action $\beta SI$, modeling the chance that too many susceptible individuals can cause a saturation effect in dynamics of the news story.  Here, $h$ is known as the ``handling time" in population ecology, or the amount of time it takes for the average ``infection" to occur.  In the context of a news story, this could be thought of as the amount of time it takes for an individual to realize the story was important enough to tweet about.  When solving the inverse problem in this case we found that the 95 percent confidence interval for $h$ was equal to $(-0.0005, 0.0034)$, which includes $h = 0$, suggesting that the infection medium (social media) is such that information transfer moves sufficiently fast so as not to require any handling time, or that this NFL story was considered ``important" roughly immediately upon the converted tweeter's read of the infectious tweet, eliciting a subsequent tweet shortly thereafter.

\section{Discussion}

In this paper we created, analyzed and fit a model for the infectiousness of the NFL's initial Deflategate story using a standard SIR epidemiological model and data from the population social networking site Twitter.  
We found that this standard SIR model fit our data quite well (see Figure \ref{fig2}), suggesting that the assumptions inherent in using such a model are reasonable.  In fact, there have been some studies (for example, \cite{McCallum2001} and \cite{Brauer2008}) suggesting that mass-action assumptions may be improperly applied in the study of actual epidemics when the underlying populations are a) too small b) not homogeneous in space enough to warrant such a simple transmission probability or c) too crowded so as to saturate infectiousness when pushed beyond a certain population size.  Populations of Twitter users, however, consist of many individuals on one webpage unimpeded physical limited, alleviating the aforementioned issues.  Thus, the simplest SIR model may, in many ways, be a better initial model for some instances of information moving through a social networking site like Twitter than it is for an epidemic moving through a real population.  

\begin{figure} 
\begin{center}
\includegraphics[scale = .5]{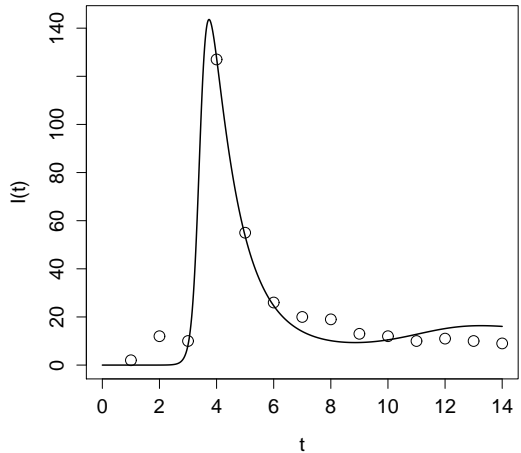}
\end{center}
\caption{The number (in thousands) of Twitter users tweeting about Hillary Clinton's announcement for candidacy for the 2016 presidential race subject to the parameter vector $\vec{\theta}$ solving the inverse problem using Topsy.com data (unfilled circles).  We used an SIR model with a sorce term for the susceptible population (i.e. $S'(t) = \Gamma - \beta S(t)I(t)$) to better fit the data. $t=0$ corresponds to April 7, 2015.}
 \label{fig3}
\end{figure}

\begin{figure} 
\begin{center}
\includegraphics[scale = .45]{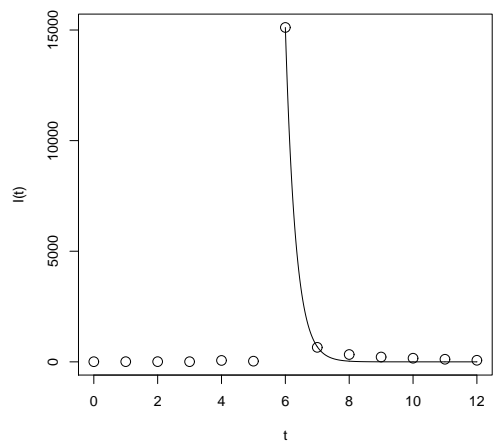}
\end{center}
\caption{The number (in thousands) of Twitter users tweeting about Freddie Gray and the Baltimore Riots subject to parameter vector $\vec{\theta}$ solving the inverse problem using Topsy.com data (unfilled circles).  Only the second half of the data (starting with $t = 6$) was fit to a simple exponential model due to the inability to fit an SIR model to all of the data. $t=0$ corresponds to April 21, 2015.}
\label{fig4}
\end{figure}

We used the parameters from the study of this inverse problem to determine how popular and persistent this news story was in terms of the composite parameter $R_0$ and the parameter $\gamma$, respectively.  We found that the average group of 1000 individuals tweeting about Deflategate were able to elicit 4.05 to 10.91 thousand new tweeters to tweet about Deflategate during the early stages of the news story, and that the average individual tweeting about the story tweeted between 2.24 and 3.53 about the story.  To put these numbers in perspective, we ran a similar analysis of two different, prominant news stories that happened since the Deflategate story:  the announcement of Hillary Clinton's presidential campaign and the stories surrounding the Baltimore riots resulting from the death of Freddie Gray.  
Interestingly, the SIR model used in this paper was not able to capture the behavior of either story, due to the almost immediate emergence of individuals tweeting about the story (see Figure \ref{fig3} and Figure \ref{fig4}).  Instead of using the traditional SIR model in these cases we used an SIR model allowing for recruitment of new susceptibles for the Clinton story and used a simple exponential model in the Freddie Gray case.  In the former we are still able to recover $\ds \mathcal{R}_0$ and $1/\gamma$ and the latter only an upper bound for $1/\gamma$.  We found that Hillary Clinton's annoucement of her presidential bid, while having more total tweets at its peak than the Deflategate story (see Figure 3), and a similar initial infectiousness ($\ds \mathcal{R}_0 \in (4.95, 7.69)$), had far less staying power.  The average tweeter was only tweeting about Mrs. Clinton's annoucement for between 0.539 and 0.7133 days.   This may be due to the fact that the election was still more than a year away, or that there were other candidates announcing their bids during the same period of time.  Be that as it may, the model for the Baltimore riot story revealed something far more disturbing.  This story, while eliciting far more tweets than the Deflategate story, saw the average tweeter tweeting about the riots for a time whose upper bound is between only 0.3227 and 0.3232 days (see Figure 4).  One explanation for this short staying time would be that, after many instances of questionable police actions over the course of the past year, many people are starting to grow fatigued by such stories.  However, one would hope for the same such fatigue over relatively meaningless NFL stories, which we have shown is not the case.  

The results in this manuscript suggest that the NFL's popularity rivals (even surpasses) that of what should be some of our nation's (and world's) biggest news stories.  Observations made in the wake of various news stories involving the intersection of a game (football) with some of our nation's most contentious issues (gambling, equality, domestic abuse, child abuse and financial literacy, to name a few) appear to coincide with these results (\cite{FlorioPFT}, \cite{Armour}, \cite{Diaz}, \cite{Stevenson}, \cite{Holmes}, \cite{Deford}).  Even Noam Chomsky, in an exerpt from \textit{The Chomsky Reader}, has pondered ``why Americans know so much about sports but so little about world affairs", arguing that sports provide the easiest way for people to concentrate each other (into fan bases) for a common (if inconsequential) goal or purpose \cite{Chomsky2014}.  This large-scale societal escape, he argues, is to combat the feeling shared by many that they simply don't have much power to solve society's real problems.  However, the rising popularity of the NFL, along with the societal stories and scandals that its members (players, coaches, execs) create, if properly leveraged, can provide society with the means to see societal change through the inevitable clash between humanity and sport.

\section{Acknowledgements} The authors would like to that the University of Wisconsin - La Crosse Eagle Apprentice program for funding the student author's involvement in this project.


\begin{thebibliography}{9}

\bibitem{Armour}
N.~Armour, N. (2014, October 22). Lack of interest in Michael Sam a good thing. Retrieved May 15, 2015, from http://www.usatoday.com/story/sports/nfl/2014/10/19/little-talk-of-michael-sam-since-joined-cowboys/17589913/

\bibitem{Avery2012}
M.~ Avery, H.T.~ Banks, K.~ Basu, Y.~ Cheng, E.~ Eager, S.~ Khasawinah, L.~Potter, K.L.~Rehm, 2012. Experimental design and inverse problems in plant biological modeling, \emph{J. Inverse Ill-Posed Probl.}, \textbf{20}(2) (2012), 169--191. 

\bibitem{Banks2014}
H.T.~Banks, S.~Hu, W.C.~Thompson, 2014. Modeling and inverse problems in the presence of uncertainty.  CRC Press, Boca Raton, FL (2014).

\bibitem{Brauer2008}
F.~Brauer, 2008. Compartment Models in Epidemiology. \emph{Mathematical Epidemiology: Lecture Notes in Mathematics}, \textbf{1945}, 2008, 19--79.

\bibitem{Cappadona2015}
B.~Cappadona, (2015, January 30). The Blizzard of 2015 Might Be Over, But Deflategate Is Forever. Retrieved May 15, from http://www.bostonmagazine.com/news/blog/2015/01/30/blizzard-is-over-but-deflategate-is-forever/

\bibitem{CCC}
C.~Castillo-Chávez, F.~Brauer, 2001. Mathematical models in population biology and epidemiology. Berlin: Springer 

\bibitem{Chomsky2014}
N~Chomsky, (2014, September 15). Why Americans Know So Much About Sports But So Little About World Affairs. Retrieved May 15, from http://www.alternet.org/noam-chomsky-why-americans-know-so-much-about-sports-so-little-about-world-affairs

\bibitem{Clark}
K.~Clark, (2015, January, 20). NFL Probes Whether Patriots Deflated Footballs Retrieved May 15, from http://search.proquest.com/docview/1646546630?accountid=9435.

\bibitem{Deford}
F.~Deford, (2014, Spetember 17).  Is the NFL too big to fail?Retrieved May 15, from http://www.npr.org/2014/09/17/348706659/is-the-nfl-too-big-to-fail 

\bibitem{Diaz}
G.~Diaz, (2014, November 30).  Ray Rice's domestic-violence story has powerful and positive impact. Retrieved May 15, from http://www.orlandosentinel.com/sports/football/os-ray-rice-george-diaz-1201-20141130-column.html

\bibitem{Doucleff}
M.~Doucleff, (2014, October 2). No, Seriously, How contagious Is Ebola?, Retrieved May 15, from http://www.npr.org/blogs/health/2014/10/02/352983774/no-seriously-how-contagious-is-ebola.


\bibitem{NFL}
Editors at the NFL. NFL Record \& Fact Book 2013 (Official National Football League Record and Fact Book). Time Home Entertainment, Inc. 688 pages.

\bibitem{PEDs}
M.~Florio, (2013, May 19). Six Seahawks have tested positive for PEDs since 2011. Retrieved May 15, from http://profootballtalk.nbcsports.com/2013/05/19/six-seahawks-have-tested-positive-for-peds-since-2011/

\bibitem{FlorioPFT}
M.~Florio, (2014, July 28).  NFL paying attention to influence of high-stakes fantasy football leagues. Retrieved May 15, from http://profootballtalk.nbcsports.com/2014/07/28/nfl-paying-attention-to-influence-of-high-stakes-fantasy-football-leagues/, July 28, 2014

\bibitem{Holling1959} 
C.S.~Holling, 1959. The components of predation as revealed by a study of small-mammal predation of the European pine sawfly. \emph{Canadian Entomologist} \textbf{91} (1959) 293–320.

\bibitem{Holmes2009} 
M.H.~Holmes, 2009.  Introduction to the foundations of applied mathematics.  Springer, New York, NY (2009).

\bibitem{Holmes}
L.~Holmes, (2012, October 2).  ESPN's `Broke' looks at the many ways athletes lose their money. Retrieved May 15, from http://www.npr.org/blogs/monkeysee/2012/10/02/162162226/espns-broke-looks-at-the-many-ways-athletes-lose-their-money

\bibitem{Kermack1927} 
W.O.~Kermack and A.G..~Kendrick, 1927. Contributions to the mathematical theory of epidemics, \emph{Proc. Roy. Soc. Er. A}, \textbf{115} (1927), 700--721.

\bibitem{McCallum2001}
H.~McCallum, N.~Barlow, J.~Hone, 2001. How should pathogen transmission be modelled?, \emph{Trends. Ecol. Evol.}, \textbf{16} (2001), 295--300.

\bibitem{R2012}
R Core Team, 2012. R: A language and environment for statistical computing. R Foundation for Statistical Computing, Vienna, Austria. ISBN 3-900051-07-0, URL http://www.R-project.org/

\bibitem{SmithThieme2011} 
H.~Smith and H.~Thieme, 2011.  Dynamical systems and population persistence.  American Mathematical Society, Providence RI.  

\bibitem{Stevenson}
A.~Stevenson, (2014, September 13).  Peterson child abuse indictment highlights NFL's violent culture.  Retrieved May 15, from http://www.commdiginews.com/sports/child-abuse-indictment-highlights-vikings-players-ties-to-violence-friendly-culture-25771/

\bibitem{Suli2003}
E.~S\"{u}li and D.F.~Mayers, 2003.  An introduction to numerical analysis.  Cambridge University Press, Cambridge, MA (2003).  

\bibitem{Twitter}
Twitter, 2015.  About Twitter.  https://about.twitter.com/company.  


\end{thebibliography}
\end{document}